\documentstyle[12pt]{article}
\def\appendixa{
\vskip 1cm
{\Large\bf Appendix A}
\vskip 1cm
\par
\setcounter{equation}{0}
\def\theequation{{\rm A.}\arabic{equation}}
}
\renewcommand{\theequation}{\thesection.\arabic{equation}}
\newcommand{\rl}{{\cal L}}
\newcommand{\be}{\begin{equation}}
\newcommand{\ee}{\end{equation}}
\newcommand{\ve}{\varepsilon}
\newcommand{\nb}[1]{\nabla _{#1}}
\newcommand{\s}{\bar s}
\newcommand{\xt}{\stackrel{\ldots}{x}}
\begin{document}
\thispagestyle{empty}
\vspace*{0.5cm}
\begin{center}
{\Large \bf Complete integrability for Lagrangian \\}
\bigskip
 {\Large \bf dependent   on acceleration in a space--time \\}
\bigskip
 {\Large \bf
 of constant curvature.
\footnote{Ricerca effettuata con i fondi  M.U.R.S.T. 40\% e 60\% art.
65 D.P.R. 382/80.}}
\vskip0.6cm
{\large  V.\ V.\ Nesterenko\footnote{{\it E-mail address:}
 Nestr@theor.jinrc.dubna.su}}
 \vskip 0.3cm
{ \it   Bogoliubov Laboratory of Theoretical Physics
 \\ Joint Institute for Nuclear Research \\
     Dubna, SU 141980 Russia}  \\
\vskip 0.3cm
and\\
\vskip 0.3cm
{\large A. \ Feoli\footnote{{\it E-mail address:}
Feoli@vaxsa.csied.unisa.it}
 \  \  \ G.\ Scarpetta\footnote{{\it E-mail address:}
Scarpetta@vaxsa.csied.unisa.it}}\\
\vskip 0.3cm
{ \it Dipartimento  di  Fisica Teorica e   s.m.s.a. --
Universit\`a di Salerno
 \\
     84081 Baronissi (SA), Italia \\
     Istituto Nazionale di Fisica Nucleare -- Sezione di Napoli \\
    International Institute for Advanced Scientific Studies
 -- Vietri sul Mare (SA)}\\
[1.4cm]

                         {\large \bf  Abstract }   \\[0.4cm]
   \end{center}
The motion equations for a
Lagrangian ${\cal L}(k_1)$, depending on the   curvature $k_1$ of  the
particle  worldline, embedded in a
space--time of constant curvature,
are  considered and reformulated in terms of the principal
curvatures.   It
is shown that for arbitrary Lagrangian function ${\cal L}(k_1)$
the general solution of the motion equations can be obtained
 by integrals. By analogy with the flat space--time case,
the constants of integration are interpreted as
the particle mass and its spin.  As examples, we  completely investigate
Lagrangians linear and quadratic in $(k_1)$ and
the model  of
relativistic  particle  with  maximal proper  acceleration, in
a space--time with constant curvature.

\bigskip

 PACS:~~14.80.Pb~~02.40.Hw
\newpage
\setcounter{footnote}0
\section{Introduction}
Recently, the  interest  to  construct  classical  models   of
relativistic  particles  with spin is again revived (see,  for
instance,  paper~[1]  and  references  therein);  between  the
motivations,  we  remember  here  the  string  approach to the
theory of elementary particles~[2, 3], the construction of the
supersymmetric  particle models~[4],  the searching for models
of particles with arbitrary fractional spin,  the  anyons~[5],
the theory of particles with maximal proper acceleration~[6].

A special class of such models are obtained, in particular, by
considering Lagrangians depending on higher order derivatives,
i.e. on the velocity, the acceleration and other higher
derivatives of the particle position vector~[7]. The
Euler-Lagrange equations in these models written in terms of
the position vector turn out to be very complicated already in
the free case. An effective method for their investigation has
been proposed in a previous paper~[8]. This geometrical
approach is based on the description of the particle world
line by its geometrical invariants, the principal curvatures,
rather than by its position vector; the closed set of
equations for principal curvatures were derived applying the
Hamilton principle, and the general solution obtained in terms
of integrals in the case of an arbitrary Lagrangian function
$\rl (k_1)$ depending on the proper acceleration of the
particle, i.e., on the curvature of the world line, $k_1$.

The investigation  of  spinning  particle   interacting   with
external  gauge and gravitational fields is an important task,
and,  as demonstrated  in  papers~[9,  10],  is  a  nontrivial
problem.  For  instance,  the  introduction  of  the  external
electromagnetic or gravitational fields into  the  model  with
Lagrangian $\rl \,=\,-\,\alpha \,k_1$ entails the violation of
the closure of the constraint algebra.

Therefore it seems to  be  worthwhile  trying  to  extend  the
geometrical  approach  proposed in ref.~[8] to the interacting
case and,  first of all,  to the case in which one takes  into
account the space-time curvature. This is the aim that will be
pursued in the present paper.  It turns out that new equations
of  motion,  in  terms  of  the  principal  curvatures  of the
particle world trajectory,  will be derived in  space-time  of
constant curvature.  Furthermore,  for an arbitrary Lagrangian
function  $\rl  (k_1)$  these   equations   can   be   exactly
integrated;  on the analogy of the flat space--time case,  one
can interpret the integration constants as mass  and  spin  of
the particle in the constant curvature space--time.

The layout of the paper is the following. In Sect. 2., after
calculating the variation  of
the proper acceleration in  a curved space-time,
we derive the equations of motion,
generated by the Lagrangian $\rl (k_1)$,
  in terms of the
principal curvatures of the world line; the basic tools used here
are the Frenet equations for the moving frame  along the curve. In
Sect. 3. the general solution to the new Euler-Lagrange equations
are  obtained in terms of integrals. The effectiveness of  this
method is illustrated by considering some examples:  Lagrangians
$\rl (k_1)$ linear in $k_1$ and the model of relativistic particle
with maximal proper acceleration.\footnote{
 These models have been investigated recently in a {\it
flat space-time}
(see references in Sect. 3.).}
In  particular, in the last model
an important inequality relating the sectional curvature of the
space--time, $G$, and the limiting acceleration of the particle,
$M_0$, is derived very easily: $M_0^2\,>\,G$. In Sect. 4.
the results obtained are shortly discussed.
In Appendix A some mathematical details are presented.

\section{Euler-Lagrange equations in terms
of the principal curvatures of the world line}

 We assume that the space-time is an arbitrary $D$-dimensional Riemannian
manifold with the metric tensor $g_{\mu \nu}(x), \;\; \mu ,\, \nu
\,=\,0,1,\ldots , D-1$ having the Lorentz signature $(+,-, \ldots, -)$.
Reduction to space-time of constant curvature will be made later. In this
manifold we shall consider the parametrized curves $x^\mu (s)$ and the
generic action
\begin{equation}
S\,=\,\int \rl (k_1)\,ds
\end{equation}
defined on them. Here $\rl $ is an {\it arbitrary} function of the first
curvature of the world line, $k_1$, i.e., of the proper acceleration of the
particle, $s$ is the arclength.\footnote{More precisely, $k_1$ is the {\it
geodesic} curvature of the world line in the Riemannian manifold.} It can
be shown [11] that any Lorentz and reparametrization invariant action with
a Lagrangian function depending on the first and on the second derivatives
of the particle position vector can be transformed into the action (2.1).

In order to shorten the formulas, the differentiation with respect to $s$
will be denoted by an overdot and the scalar product generated by the
metric tensor $g_{\mu \nu }(x)$ will be denoted by $<\ldots,\, \ldots >$.
We shall also use the notion of {\it the covariant differentiation along
the vector field} $A(x):\;\nabla _A$. In these notations one has~[12, 13]
\begin{equation}
<\dot x,\,\dot x >\,=\,1,
\end{equation}
\begin{equation}
k_1^2\,=\,-\, <\nabla _{\dot x}\,\dot x, \,\nabla _{\dot x}\,\dot x >,
\end{equation}
where $\dot x $ is the tangent vector field associated with the particle
trajectory $x^\mu (s)$. In the usual notations $\nabla _{\dot x}\,\dot x$
is nothing else  but
$$
\frac{d^2x^\mu}{ds^2}\,+\,\Gamma ^\mu _{\nu \rho}\frac{dx^\nu }{ds}\,
\frac{dx^\rho }{ds},
$$
where $\Gamma ^\mu _{\nu \rho}$ are the Christoffel symbols for the metric
$g_{\mu \nu}(x)$.

To obtain the equations of motion we, as usually, equate to zero the
variation of the action (2.1)
\begin{equation}
\delta S\,=\,\delta S_1\,+\,\delta S_2\,=\,0,
\end{equation}
where
\begin{equation}
\delta S_1\,=\,\int ds\,\delta \rl(k_1)\,=\,\int ds\,\rl '(k_1)\,\delta k_1,
\end{equation}
\begin{equation}
\delta S_2\,=\, \int \rl (k_1)\,\delta(ds).
\end{equation}
The prime of the Lagrangian function $\rl$
denotes the differentiation with respect to its argument $k_1$.

Under  variation, as usually [13],
the position vector of the curve is treated as a function of two variables,
$x^\mu (s,\,\xi)$, with the condition $x^\mu (s,\,0)\,=\,x^\mu(s)$.
Associated with such a variation is the vector  field
\be
\xi ^\mu(s)\,=\,\left .\frac{\partial x^\mu (s,\,\xi)}{\partial \xi}\,
\right |_{\xi \,=\,0}
\ee
defined along the curve $x^\mu(s)$. The variation of an arbitrary
 function of $x$, $F(x)$, is given by
\be
\delta F\,=\,\left .\frac{\partial F(x(s,\,\xi))}{\partial \xi}\,
\right |_{\xi = 0}\delta \xi\,=\,\xi ^\mu\frac{\partial
 F}{\partial x^\mu }\,\delta \xi\,{,}
\ee
or in componentless notation
$$
\frac{\partial F}{\partial \xi}\,=\,\xi \circ F\,{.}
$$

In order to calculate the variations, it is convenient to use the Frenet
frame associated with the world curve $x^\mu (s)$:
\begin{equation}
\nu _\alpha (s),\;\; \alpha \,=\,0,1, \ldots, D-1,
\end{equation}
$$
\nu _0\,=\,\dot x, \;\;<\nu_\alpha, \,\nu _\beta >\,
=\,\eta_{\alpha \beta}\,{,}
$$
\begin{equation}
\eta _{\alpha \beta}\,=\,\mbox{diag}\,(1,-1,\ldots,-1),\quad
\alpha,\,\beta\,=\,0,1,\ldots,D-1
\end{equation}
and the Frenet equations describing the motion of this basis along the world
line~[12]
\begin{equation}
\nabla _{\dot x}\, \nu_\alpha \,=\,\omega _\alpha {}^\beta {}\,\nu _\beta,
\quad \omega _{\alpha \beta}\,+\,\omega _{\beta \alpha}\,=\,0\,{.}
\end{equation}
The rising and lowering the frame indexes
$\alpha,\, \beta,\, \gamma,\, \ldots $
are made by  the
constant diagonal tensor $\eta _{\alpha \beta}$ (2.10). Nonzero
elements of the matrix $\omega $ are determined by the
principal curvatures of
the world line
\begin{equation}
\omega _{\alpha, \alpha + 1}\,=\,-\, \omega _{\alpha +1,
\alpha}\,=\,k_{\alpha +
1}(s), \quad \alpha \,=\,0,\,1,\,\ldots ,\,D-2.
\end{equation}
In the Frenet basis, the variation $\delta x^\mu $ will be defined by the
expansion
\begin{equation}
\delta x^\mu (s) \,=\,\xi ^\mu (s)\,\delta \xi\,=\,
\varepsilon ^\alpha (s)\,\nu _\alpha ^\mu (s)\,{,}
\end{equation}
$$
\mu \,=\,0,\,1,\,\ldots ,\,D\,-\,1, \quad \alpha \,=\, 0,\,1,\,\ldots, \,
D-1,
$$
where $\ve ^\alpha (s)$ are arbitrary functions.

For the variation $\delta S_2$ we have
\be
\delta S_2\,=\,\int \rl (k_1)\,\dot x ^\nu g_{\mu \nu}\,d
(\delta x^\mu )\,+
\,\frac{1}{2}\, \int ds \,\rl \,\dot x^\mu\,\dot x^\nu \,\frac{\partial
 g_{\mu \nu }}{\partial x^\lambda }\,\delta x^\lambda\,{.}
\ee
Here we have used the commutativity of the symbols $\delta $ and $d$.
Partially  integrating in the first term
and dropping the terms outside the
integral, as it is  always done when deriving the equations of motion from
the Hamiltonian principle, we arrive at the formula
\be
\delta S_2\,=\,-\,\int \rl '\,\dot k_1<\dot x,\,\delta x>ds\,-\,
\int \rl <\nabla _{\dot x}\dot x,\,\delta x>ds\,{.}
\ee
Taking into account the Frenet equations (2.11) and the expansion (2.13), the
variation $\delta S_2$ acquires the form
\be
\delta S_2\,=\,-\,\int \rl '(k_1)\,\dot k_1\,\ve ^0(s)\,ds\,-\,
\int \rl (k_1)\,k_1(s)\,\ve ^1(s)\,ds\,{.}
\ee

Let us now turn to the calculation of the variation $\delta S_1$ defined by
(2.5). At this end we firstly calculate the variation of the first
curvature $k_1(s)$ in terms of $\delta x^\mu$. For two arbitrary vector
fields $A(x)$ and $B(x)$, the following equality holds:
\be
\frac{\partial }{\partial x^\mu }<A,\,B>\,=\,<\nabla _\mu \,A,\,B>\,+
\,<A,\,\nabla _\mu \,B>
\ee
 Making use of (2.3), (2.8) and (2.17), we obtain the following
expression for the variation of the curvature $k_1(s)$
\be
\delta k_1^2\,=\,2\,k_1\,\delta k_1\,=\,-\,2<\nabla_{\dot x}\,\dot x,\,
\nabla _\xi \nabla _{\dot x}\,\dot x>\delta \xi\,{.}
\ee

Commutator of two covariant derivatives can be expressed in terms of the
Riemann curvature tensor. For three vector fields $A,\;B$, and $C$
we have the  first structure equation~[13]
\be
\nb {A}\,\nb {B}\,C\,-\,\nb {B}\,\nb {A}\,C\,-\,\nb {[A,\,B]}\, C\,=\,
R(A,\,B)\,C
\,{,}
\ee
where $[A,\,B]$ is the commutator of the vector fields
$A$ and $B$
\be
[A,\,B]\,=\,\left( A^\nu \frac{\partial B^\mu }{\partial x^\nu }\,-\, B^\nu
\frac{\partial A^\mu }{\partial x^\nu}\right )
\,\frac {\partial }{\partial x^ \mu}\,{.}
\ee
This enables us to substitute $\nb {\xi }\,\nb {\dot x}\,\dot x$ in
(2.18) by the following expression
\be
\nb{\xi}\,\nb{\dot x}\,\dot x\,=\,
\nb {\dot x}\,\nb {\xi }\,\dot x\,+\, \nb {[\xi,\,\dot x]}\,\dot x\,+\,
R(\xi,\, \dot x)\,\dot x\,{.}
\ee
The torsion of the space-time is assumed to be zero. Therefore the second
structure equation reads~[13]
\be
\nb {\xi }\dot x
\,-\,\nb {\dot x}\,\xi
\,-\,[\xi,\,\dot x]\,=\,0\,{.}
\ee
In Appendix A it is shown that
\be
[\xi,\,\dot x]\,=\,-\,<\dot x,\,\nb{\dot x}\,\xi > \dot x\,{.}
\ee
Hence,  eq.~(2.21) acquires the form\footnote{Operator $\nb{\dot x}$
acting on the scalar function reduces to the usual differentiation
$d/ds$.}
\begin{eqnarray}
\nb {\xi}\,\nb{\dot x}\,\dot x &=& \nb {\dot x}\,\nb {\dot x} \,\xi \,-\,
2<\dot x,\,\nb{\dot x}\,\xi >\nb{\dot x}\,\dot x\,- \nonumber \\
&& -\,\frac{d}{ds}\,(<\dot x,\,\nb{\dot x}\,\xi>)\,\dot x\,+\,
R(\xi,\, \dot x)\,\dot x
\,{.}
\end{eqnarray}
Substituting (2.24) into (2.18) and taking into
account that $<\nb{\dot x}\,\dot x ,\,\dot x>\,=\,0$, we
can write
\begin{eqnarray}
k_1\,\delta k_1&=&-\,\delta \xi\,\left \{ <\nb{\dot x}\,\dot x,\,
\nb{\dot x}\,\nb{\dot x}\, \xi>\,+ \right . \nonumber \\
&& \left .+\,2\,k^2_1<\dot x,\,
\nb{\dot x}\,\xi>\,+\,<\nb{\dot x}\,\dot x,\,
R(\xi,\,\dot x)\,\dot x>\right \}\,{.}
\end{eqnarray}
It should be noted here
that $\delta \xi$ in eq.~(2.25) is a variation
of the independent variable which enters as an argument of the
 function $x^\mu (s,\,\xi)$. Hence, $\delta \xi$ obviously commutes
 with the differentiation operator $\nb{\dot x}$. Now we expand
$\xi ^\mu\,\delta  \xi$ in (2.25) according to (2.13) and use the Frenet
 equations (2.11). As a result, the variation $\delta k_1(s)$ can be
represented~as
\begin{eqnarray}
\delta k_1 &=&\ve ^0\,\dot k_1\,-\,\ddot \ve ^1\,+\,\ve ^1\,\left (
k_1^2\,+\,k_2^2\right )\,-\,2\,\dot \ve ^2\,k_2\,-\,\ve^2\,
\dot k_2 \,- \nonumber \\
&& -\, \ve ^3\,k_2\,k_3\,+\,\sum _{\alpha\,=\,0}^{D-1}\ve^\alpha (s)
<\nu_1,\,R(\nu_\alpha,\,\nu _0)\nu _0 >\,{.}
\end{eqnarray}
In the component notations the last term in (2.26) is written as
$$
\sum_{\alpha \,=\,0}^{D-1}\ve ^\alpha (s)\,
<\nu_1,\,R(\nu _\alpha ,\,\nu_0)\,
\nu _0>\,=\,\ve ^\alpha (s)\, R_{\mu \nu \rho }{}{}{}^\sigma
\nu_\alpha ^\mu \,\nu_0^\nu \, \nu_0^\rho \,\nu_{1 \sigma}\,{.}
$$
Thus for an arbitrary Riemann curvature tensor $R_{\mu \nu \rho}{}{}{}
^\sigma $,  the variation $\delta k_1$ is expressed not only in terms
of the principal curvatures of the world line, $k_j(s), \;\;j\,=\,1,\,
2,\,3$ but it also depends on the normals $\nu _\alpha (s)
, \;\;\alpha \,=\,0,\,1,\,\ldots,\,D-1$ to the curve. As a consequence,
for an arbitrary curvature tensor $R_{\mu \nu \rho}{}{}{}^\sigma$,
we cannot derive a
closed set of equations of motion containing only $k_j(s)$;
however, for special Riemannian manifolds, dependence on the Frenet
frame in (2.26) may disappear. For example, the space-time of
constant sectional
curvature $G$ has the Riemann tensor defined by~[13]
$$
R_{\mu \nu \lambda \rho}\,=\,G\,(g_{\mu \rho}\,g_{\nu \lambda}\,-
\,g_{\mu \lambda}\,g_{\nu \rho})\,{.}
$$
Scalar curvature $R\,=\,g^{\mu \rho}\,g^{\nu \lambda}\,R_{\mu \nu
\lambda \rho}$ is related to the sectional curvature $G$ in the
following way $R\,=\,D\,(D\,-\,1)\,G$, where $D$ is the dimension
of the space-time. Now we have
\be
\sum_{\alpha\,=\,0}^{D-1}\ve ^\alpha (s)<\nu_1,\,R(\nu_\alpha,
\,\nu_0)\,\nu_0>\,=\,-\,G\,\ve^1(s)\,{.}
\ee
Taking into account (2.26) and (2.27), the variation $\delta S_1$
can be represented in
the form
$$
\delta S_1
\,=\,\int ds \left \{\rl '(k_1)\,\dot k_1\,\ve^0(s)\,+\right .
$$
\be
 +\, \left [ (k_1^2\,+\,k_2^2) \,\rl '(k_1)\,-\,\frac{d^2}{ds^2}
\, \left (\rl '(k_1)
\right )\,-\, \rl '(k_1)\,G
\right ]\,\ve ^1(s)
\ee
$$
 \left .+\, \left [2\,\frac{d}{ds}\left (\rl '(k_1)\,k_2
\right )\,-\,\dot k_2\, \rl '(k_1)
\right ]\, \ve ^2(s)\,-\,\rl '(k_1)\,k_2\,k_3\,\ve ^3(s)\,
\right \}\,{.}
$$
Summing eqs.~(2.14) and (2.28), we obtain the  set of three equations
for principal  curvatures $k_1,\,k_2$, and $k_3$ (terms containing
$\ve ^0(s)$ in $\delta S_1$ and in $\delta S_2$ are canceled):
\begin{eqnarray}
\frac{d^2}{ds^2}\left( \rl '(k_1)\right) &=& (k_1^2\,+\,k_2^2\,-\,G)\,
\rl '(k_1)\,-\,k_1\,\rl(k_1)
\,{,} \\
2\,\frac{d}{ds}\left( \rl '(k_1)\,k_2
\right)&=& \dot k_2\,\rl '(k_1)\,{,} \\
\rl '(k_1)\,k_2\,k_3 &=& 0\,{.}
\end{eqnarray}
It is remarkable that the constant sectional
 curvature of the space-time, $G$, enters
only in eq.~(2.29). Equations (2.30) and (2.31) remain the same
as  in the flat space-time [8].

In order to satisfy eq.~(2.31) we put $k_3(s)\,=\,0$.
Then all the higher curvatures will  vanish too~[14]. Thus, for arbitrary
$D$ we have
\be
k_n(s)\,=\,0,\quad n\,=\,3,\,4,\,\ldots,\,D-1\,{.}
\ee

Equation (2.30) can be integrated
\be
\left (\rl '(k_1)
\right )^2\,k_2\,=\, C\,{,}
\ee
where $C$ is an integration constant. Taking into account (2.33),  we
remain with
one nonlinear equation of the second order for the curvature $k_1(s)$
\be
\frac{d^2}{ds^2}\left (\rl '(k_1)
\right )\,=\, \left ( k_1^2\,+\, \frac {C^2}{\left (\rl '(k_1)
\right )^4}\,-\,G
\right )\,\rl '(k_1)\,-\,k_1\,\rl (k_1)\,{.}
\ee
After integrating this last equation, we can reconstruct all the principal
curvatures of the world line $k_i(s), \;\;i\,=\,1,\,2,\, \ldots ,\,D-1$.
 From the classical differential geometry~[12], it is well known that the
principal curvatures of a world line in a background of constant sectional
curvature determine this curve up to its transformations as a whole, which
are given by the symmetry group of the enveloping space (in the case under
consideration it is the $SO(1,\,D-1)$ group). Obviously, this specification
of the world line should take into account all the essential physical
properties of the model in question.

\section{Exact integrability of the Euler-Lagrange equations for
principal curvatures. Examples}
\setcounter{equation}{0}

The first integral for equation (2.34) can be found directly. As a result
the problem of constructing the general solution to this equation is
reduced to quadratures. In Ref.~[8] in the case of a flat space--time, this
integral has been derived by investigating the equations of motion,
generated by action (2.1), in terms of the position vector $x^\mu$ of the
world line. To accomplish similar calculations in the case under
consideration is a rather complicated task. Nevertheless knowing the
integral for eq.~(2.34) at $G\,=\,0$ one can construct such an integral at
$G\,\neq \,0$ too. Really, by a direct differentiation one can convinced
oneself that
 the expression
\be
M^2\,=\,\rl ^2\,-\,\left ( \frac{d}{ds}\rl '\right )^2\,-\,
2\,\rl \,\rl ' \,k_1\,+\,\left ( \rl '\right )^2\,k_1^2\,-\,\frac{C^2}{
(\rl ')^2}\,-\,G\,\left (\rl '
\right )^2\,{,}
\ee
 $M^2$ being an integration constant, is the
first integral of the eq.~(2.34) if $\dot k_1\rl ''\,\neq \,0$. For
Lagrangians linear in $k_1$, eqs.~(3.1) and (2.34) should be treated
as independent ones, because in this case the differentiation
of (3.1) gives identically zero. From (3.1) we obtain
\be
\frac{dk_1}{ds}\,=\,\pm \sqrt{f(k_1)}\,{,}
\ee
where
\be
f(k_1)\,=\,\frac{1}{(\rl '')^2}\left \{\rl ^2\,-\,2\,\rl \,\rl '\,k_1\,+\,
(\rl ')^2\,(k_1^2\,-\,G)\,-\,\frac{C^2}{(\rl ')^2}\,-\,M^2
\right \}\,{.}
\ee
Integration of eq.~(3.2) yields
\be
\int\limits _{k_1(s_0)}^{k_1(s)}\frac{dx}{\sqrt{f(x)}}\,=\,\pm\,
(s\,-\,s_0)\,{.}
\ee
Thus, making use of (3.4), one can obtain the first curvature $k_1(s)$
of the world trajectory for any given Lagrangian function $\rl (k_1)$.
Then eq.~(2.33) enables one to find the torsion $k_2(s)$ of
the world line, all remaining principal  curvatures being equal to zero
identically. Hence, the problem of solving the equations
of motion  reduces to doing
the integral (3.4).

As it was shown in Ref.~[8] for a flat space-time, the constant of
integration $M^2$ turns out to be the mass squared of the
particle\footnote{In  particle models with  action containing
higher derivatives, the particle mass and  spin  are, at
classical level, simply the {\it integrals of motion}, whose
values  are determined by the initial conditions
for the corresponding Euler-Lagrange equations.
Therefore these integrals may,
in principle, acquire {\it arbitrary values}. Only upon quantization,
in some models of this kind one can obtain {\it discrete
values} for $M^2$ and
$S$ [7, 15].} and the second integration constant, $C$,
determines the partcle
spin  $S$
\be
S^2\,=\,\frac{C^2}{|M^2|}\,=\,\frac{k_2^2\,(\rl '(k_1))^4}{|M^2|}\,{.}
\ee
Here the absolute value of $M^2$ is taken, because in general these
models  may also have tachyonic solutions.

For a curved space-time and in particular for space-time with
constant curvature there is no unique prescription for defining
the particle mass
 and  spin. Therefore on  the analogy of the
flat space-time case, one can treat eqs.~(3.1) and (3.5)
as  definitions of the  particle mass and spin in the case of particles
 with
action (2.1), moving in   constant curvature space--time. Applying
these formulas to the usual scalar particle with  action
$$
S_0\,=\,-\,m\int ds
$$
one obtains $M^2\,=\,m^2$ and $S\,=\,0$, i.e., mass and spin of the
particle remain the same as in the flat space-time.

It is worthwhile to note that
our equations (2.29)--(2.31), as well as (3.1) and (3.5),
solve the variational problem (2.1) (2.4)
in a class of {\it regular curves} $x^\mu(s)$, i.e., in the class in which
the standard Frenet frame
 (2.9), (2.10) can be associated with each point of the world line.

Let us now analyse some examples.
We begin with
the Plyushchay model of the massless spinning particle defined by
the Lagrangian~[15, 16]
\be
\rl \,=\,-\,\alpha \,k_1(s)\,{.}
\ee
In the Minkowski space-time the principal curvatures are:
$$
k_1(s) \mbox{ is an arbitrary function of $s$ and}
$$
\be
k_2(s)\,=\,k_3(s)\,=\,\ldots =\, k_{D-1}(s)\,=\,0\,{.}
\ee

     The solutions    obtained in ~[15], in
form of  helical curves, have just zero torsion and constant
curvature. The  last  condition  can be treated as a consequence of the
 gauge fixing. The essential point is that the conditions
     $k_2(s)\,=\,0$ and    $k_1(s)\,=\,$constant   imply
superlight velocities.  Really,  from  the   Frenet   equations
(2.11), which  are  at  the  same  time the definitions of the
principal curvatures $k_j$, we obtain when $k_2\,=\,0$
     \begin{equation}
     {\xt}^2\,=\,-\,{\dot k}_1^2\,+\,k_1^4\,\dot x^2\,{.}
     \end{equation}
 From physical considerations,  the vectors  $\ddot x $  and  $\xt $
should be   treated  naturally   as  space-like vectors (this follows
immediately in the gauge in which the evolution parameter is $x^0$).
In view of this,  one deduces from (3.8) that the vector $\dot
x^\mu$ must be space--like\footnote{In Ref.~[8],  as  well
as in the present paper,  it has been
assumed that the world lines  must  be  time-like  $(\nu
_0^2\,=\,1)$. However,  taking into account the remark made before eqs.\
(2.29)--(2.31), it is obvious that these equations remain
the same in the case of space-like curves too.} when $\dot k_1\,=\,0$.

In the case of constant curvature space-time,  the solutions
to eqs.~(2.29)--(2.31) for the Lagrangian (3.7) are
$$
k_1(s) \mbox{ remains arbitrary and}
$$
\be
k_2^2\,=\,G,\;\;k_3(s)\,=\,k_4(s)\,=\ldots =\,k_{D-1}(s)\,=\,0\,{.}
\ee
Hence, the theory will be consistent only for space-time with constant
curvature $G$. Probably, this point is related to the unclosure of the
constraints algebra which has been recovered in Ref.~[10] by considering
the Hamiltonian formalism for this model in a curved space-time.

Using eqs.~(3.1) and (3.5) proposed by us for the particle mass and spin
in space-time of  constant sectional curvature
$G$, we obtain for the Lagrangian~(3.7)
\be
M^2\,=\,-2\,\alpha ^2 G,
\ee
\be
S^2\,=\,\alpha ^2/2\,{.}
\ee
As it was noted above, the consistency of the model requires $G\,\geq \,0$.
 Therefore, from eq.~(3.10) it follows that all the solutions
in this model are tachyonic with $M^2 \,<\,0$. From (3.10) and (3.11)
we obtain the relation between $M^2$ and $S^2$
\be
M^2\,=\,-\,4\,G\,S^2\,{.}
\ee
In quantum case the parameter $\alpha $, according to (3.11), becomes
discrete
\begin{equation}
\alpha^2=2\,S\,(S+D-1),\qquad S=0,1,\ldots\,{.}
\end{equation}

Let us now consider the model defined by the Lagrangian
\begin{equation}
\rl (k_1)=\,-\,m\,-\,\alpha \,k_1 (s).
\end{equation}
In a flat space this model has been examined in papers~[7, 17].
 From the equations of motion (2.29)--(2.31)
it follows that in the  space-time of  constant
curvature the principal curvatures $k_1$ and $k_2$ are
constants obeying the relation
\begin{equation}
\alpha \,k^2_2\,=\,\alpha \,G\,+\,m\, k_1\ .
\end{equation}
All   the other curvatures are equal to zero.
Equations (3.1) and (3.5) give for the mass and the spin of the particle
\begin{eqnarray}
M^2&=& {m^2-\alpha^2 \,G\over 1\,+\,\alpha^{-2} S^2}\ ,\\& &\\
S^2&=&{\alpha^4\, k^2_2\over M^2}\,{.}
\end{eqnarray}

Thus, the curvatures $k_1$ and $k_2$ are ultimately expressed in
terms of $M^2$ and $S^2$. Equation
(3.16) is the spin-mass relation.
 As compared with the flat space-time there appears an additional
term $-\alpha^2 \,G$ in eq.~(3.16).

Let us consider the rigid relativistic particle with  Lagrangian
\begin{equation}
\rl\,=\,-\,m-\alpha \,k^2_1\,{.}
\end{equation}
In flat space-time this model has been investigated in papers [11, 17, 21].
In space-time of  constant curvature the
equations of motion for $k_j (s)$ read
\begin{eqnarray}
&&2\,\ddot k_1\,+\,
k_1^3\,-\,2\,k_1\,k_2^2\,+\,
2\,G\,k_1\,+\,\alpha^{-1} k_1\,m\,=\,0\ ,\nonumber\\
&&16\, \alpha^4\,k_1^4 \,k^2_2\,=\, S^2\,M^2\ ,\\
&&k_3\,=\,k_4\,=\,\ldots \,=\, k_{D-1}=0\ .\nonumber
\end{eqnarray}
Applying general formula
(3.4), the solution to (3.19) can be expressed in terms
of  elliptic integrals, in  complete analogy
with previous investigations of this model in flat space-time~[14, 21].
Therefore we do not present here the
corresponding formulas.

And finally we consider the model of  relativistic
particle with maximal proper acceleration~[6]
\begin{equation}
\rl (k_1)=-\mu_0\sqrt{M_0^2-k_1^2}\ ,
\end{equation}
where $\mu_0 = m/M_0$ and $M_0$ is the upper value of the
proper acceleration of the  particle.\footnote{In  ref.\  [23]
this Lagrangian has been investigated in another context.}
Obviously, for physical applications it
is interesting to investigate the behaviour of
$k_1(s)$
near the boundary $M_0$. In this region equations (3.2)--(3.4) give
\begin{equation}
\int \limits ^{k_1(\s)}_{k_1(\s_0)}
\frac{dk}{\sqrt{1-k^2}}\,=\,\pm\, \sqrt{1\,-\,g}
\left(\s\,-\,\s_0\right)\ ,
\end{equation}
where $k(\s)$ is a dimensionless acceleration, $k^2=k^2_1/M_0$, $\s =s\,
M_0$ is the dimensionless
arclength, $g$ stands for the ratio of the sectional
curvature of the background space--time to the squared maximal
acceleration $M_0^2$, $g=G/M_0^2$. From (3.21) it follows that
the model under consideration is consistent
only when\footnote{The sectional curvature of the space-time,
 $G$, is supposed to be positive. The
space-time with $G<0$ has rather unusual properties; for example, the
time-like closed geodesic curves
may  exist there~[22].}
\begin{equation}
M_0^2\,>\,G\ .
\end{equation}
Integration in (3.21) gives
\begin{equation}
k^2(\s) \,=\,\tanh^2\left[\sqrt{1\,-\,g} (\s\,-\,
\s_0)\right]{,}\quad  \s\,\to\,\pm\,\infty\ .
\end{equation}
Hence, the restriction
$$
k^2(\s)\,<\,1
$$
always holds.

\section{Conclusion}

The geometrical  approach  for  investigating  the models with
action (2.1) in space-time of  constant  curvature,  proposed
here, enables us, without complicated calculations, to reveal
the basic  important  particularities  in  this  dynamics.  As well
known,  the  reparametrization invariant actions,  like (2.1),
give rise to constrained  dynamics~[7],  both  in  Lagrangian  and
Hamiltonian settings.  The analysis of the constraints and the
choice of an appropriate gauge fixing condition turn  out  to
be  a  rather  complicated  task.  In  this  regard,  it  is
remarkable that our approach allows us to avoid this  problem.
Besides,  in  the framework of the geometrical treatment, there
appears a possibility for introducing in a consistent way  the
definition  of  particle  mass  and   spin    in
space-time of  constant  curvature  as  special  integrals  of
motion.  Certainly,  it is
interesting  to  elucidate  the  relationship  between   these
definitions and other approaches to this problem.

\bigskip

\noindent{\Large \bf Acknowledgments}

\bigskip
This paper has been accomplished during the stay of one of the authors,
V.V.N., at the Department of  Theoretical  Physics
in Salerno University. Taking this opportunity he would like to thank
Prof.\ G.~Scarpetta and
his colleagues for their warm hospitality.
     Some topics  of  this paper have been discussed with Dr.\
M.S.~Plyushchay. V.V.N.\ is thankful him for this.
Partial support from
 the Russian Foundation for Fundamental Research  under
project 93-02-3972  is  gratefully acknowledged.

\appendixa
\setcounter{equation}{0}
By making use of  the definition of the commutator (or the Lie bracket)
of two vector fields (2.20), we can write
$$
[\xi,\,\dot x]\,=\,\left ( \xi ^\nu\,\frac{\partial
\dot x^\mu}{\partial x^\nu}\,-\,\dot x^\nu\,\frac{\partial
\xi ^\mu}{\partial x^\nu}
\right)\,\frac{\partial}{\partial x^\mu} \,=
$$
\be
\,=\,\frac{\partial}{\partial \xi}\,\frac{d}{ds}\,-\,\frac{d}{ds}\,
\frac{\partial}{\partial \xi}\,=\,-\,\frac{1}{ds}\,\left(
\frac{\partial}{\partial \xi}\,ds
\right )\,\frac{d}{ds}\,{.}
\ee
Further we have
\be
\frac{\partial}{\partial \xi}\,ds\,=\,
\frac{\partial}{\partial \xi}\sqrt{dx^\mu\,dx^\nu\,g_{\mu \nu}(x)}\,=\,
\left(
\dot x^\mu \,\dot \xi ^\nu \,g_{\mu \nu}\,+\,\frac {1}{2}\,
\frac{\partial g_{\mu \nu}}{\partial x^\rho}\,\xi ^\rho \,\dot x^\mu
\,\dot x^\nu
\right )\,ds\,{.}
\ee
In eq.~(A.2) $\partial g_{\mu \nu}/\partial x^\rho$ can be substituted
by
$$
\frac{\partial g_{\mu \nu}}{\partial x^\rho}\,+\,
\frac {\partial g_{\rho \mu}}{\partial x^\nu}\,-\,\frac
{\partial g_{\nu \rho}}{\partial x^\mu}
$$
because the additional terms are cancelled.
As a result, eq.~(A.2) transforms to
\be
\frac{\partial}{\partial \xi}\,ds\,=\,
\dot x^\lambda\,g_{\lambda \mu}\left (\dot \xi ^\mu \,+\,\Gamma
^\mu_{\nu \rho}\,\xi ^\rho\,\dot x^\nu
\right )\,ds\,
=\,<\dot x,\,\nb{\dot x}\,\xi>ds\,{.}
\ee
Finally we obtain
\be
[\xi,\,\dot x]\,=\,-\,<\dot x,\,\nb{\dot x}\,\xi>\frac{d}{ds}\,=\,-
<\dot x,\,\nb{\dot x}\,\xi>\dot x\,{,}
\ee
where $\dot x$ should be treated as a vector field.

\end{document}